\documentclass[12pt,a4paper]{article}
\usepackage{amsmath}
\usepackage[dvips]{graphicx}
\input epsf
\usepackage{times}
\begin{document}
\begin{titlepage}
\title{Directed flow as  effect of transient matter
rotation in hadron and nucleus collisions}
\author{S.M. Troshin, N.E. Tyurin\\[1ex]
\small  \it Institute for High Energy Physics,\\
\small  \it Protvino, Moscow Region, 142281, Russia}
\normalsize
\date{}
\maketitle

\begin{abstract}
 We discuss
directed flow $v_1$, observable introduced for description of nucleus collisions.
We consider its possible origin  in hadronic  reactions as a result of rotation
of the transient matter and trace analogy with nucleus collisions.
\end{abstract}
\end{titlepage}
\setcounter{page}{2}

\section*{Introduction}
Multiparticle production in hadron and nucleus collisions and
corresponding observables  provide  a clue to the mechanisms of
confinement and hadronization. Discovery of the deconfined state
of matter has been announced  by the four major experiments at RHIC
\cite{rhic}. Despite the highest values of energy and density have
been reached, a genuine quark-gluon plasma QGP (gas of the free
current quarks and gluons) was not found \footnote{It is  to be
noted here that confinement due to causality principle might exclude the
 very existence of QGP defined
that way \cite{mrowch}.}. The deconfined state
reveals the properties of the perfect liquid, being strongly
interacting collective state and therefore it was labelled as sQGP
\cite{denteria}.  The nature of new form of
matter discovered is not known and  variety of models has been proposed to
treat its properties \cite{models}. The importance of the experimental discoveries
at RHIC
is that the matter is  strongly correlated
and reveals high degree of the coherence when it is
 well beyond the critical values of density and
temperature.

Important tools in the studies of the nature of the new form of matter are the
anisotropic flows which are the quantitative characteristics of the collective
motion of the produced hadrons in the nuclear interactions. With their measurements
one can obtain a valuable information on  the early stages of reactions and observe
signals of QGP formation \cite{s1,s2,s3,s4,s5,s6,s7,s8,s9}.

The measurements of anisotropic  flows and constituent quark scaling
  demonstrated  an importance of the constituent quarks
 \cite{volosh} and their role as
effective degrees of freedom of the newly discovered state of matter.

The recent review paper \cite{weiner}  provides an emphasis on the historic aspects of the QGP
searches. The important conclusion made in
this work is that the  deconfined state of matter has
being observed in hadronic reactions during a long time and it would be
interesting to study collective properties of transient state in
reactions with hadrons and nuclei simultaneously. Thus, keeping in mind
that the dynamics of strong interactions is the same for the hadron
and nucleus collisions, the following question naturally arises, namely,
what, if anything, the recent discoveries at RHIC  implement
 for hadronic interactions?

  In this note we try to address in a model way  one aspect of this
broad  problem, i.e.
we  discuss the role of the coherent rotation
of the transient matter in hadron collisions as the origin of the directed
flow in these reactions and stress that  behavior of collective observable $v_1$
in hadronic and nuclear reactions would be similar.

The already mentioned experimental probes  of collective dynamics in
$AA$ interactions \cite{voloshin,molnar}, the momentum anisotropies
$v_n$  are defined
by means of the Fourier expansion of the transverse momentum spectrum
over the momentum azimuthal angle $\phi$.
The  angle $\phi$ is
the  angle of the detected particle transverse momentum with respect to the reaction
plane spanned by the collision axis $z$ and the impact parameter vector $\mathbf b$
 directed along the $x$ axis. Thus, the anisotropic flows are  the
 azimuthal correlations with the reaction plane.
In particular, the directed flow is defined as
\begin{equation}\label{dirfl}
v_1(p_\perp)\equiv \langle \cos \phi \rangle_{p_\perp}=
\langle {p_x}/{p_\perp}\rangle =
\langle {\hat{\mathbf  b}\cdot {\mathbf  p}_\perp} /{p_\perp}\rangle
\end{equation}

From Eq. (\ref{dirfl}) it is evident that this observable can be used for
studies of multiparticle production dynamics in hadronic collisions provided
that impact
parameter $\mathbf  b$ is fixed.
Therefore we   discuss hereabout   directed
flow $v_1$ in hadron collisions at
 fixed impact parameters and refer after that those considerations to the
collisions of nuclei
 proceeding from  the same nature  of the transient states in both cases.
  It is reasonable to proceed that way
 in the framework
of the constituent quark model picture   where hadron structure  looks  similar to the
structure of light nuclei. In particular, we  amend the model \cite{csn,jpg} developed for hadron interactions
 (based on the chiral quark
model ideas)  and consider the effect of collective rotation of a quark matter in the overlap region.
  We formulate a hypothesis on connection
  of the strongly interacting transient matter rotation with the directed flow generation.

The outline of the paper is the following.
In section 1 the dynamics of hadronic interaction, nature of transient state and effective
degrees of freedom of the transient state are treated. Section 2 is devoted to the specific
role of the transient matter rotation as an origin of the directed flow in hadronic reactions.
The energy, rapidity and
transverse momentum dependencies are discussed there. Section 3 considers similarity and
difference of nuclei with hadron collisions and discussion of the recent experimental measurements
 of the directed flow at RHIC.
In the final section we discuss  possible  relation of the directed flow with
 the nature of transient matter, i.e. is the matter strongly interacting or
a weakly interacting  one at the LHC energies and provide  a conclusion.

\section{Effective degrees of freedom and transient state of matter in hadron collisions}

We would like to point out the possibility that
the origin of the transient state and its dynamics along with hadron structure can be related to
the mechanism of spontaneous chiral symmetry breaking ($\chi$SB) in QCD \cite{bjorken},
 which  leads
to the generation of quark masses and appearance of quark condensates. This mechanism describes
transition of the current into  constituent quarks.
The  gluon field is considered to be responsible for providing quarks with
  masses and its internal structure through the instanton
  mechanism of the spontaneous chiral symmetry breaking.
Massive  constituent quarks appear  as quasiparticles, i.e. current quarks and
the surrounding  clouds  of quark--antiquark pairs which consist of a
mixture of quarks of the different flavors.  Quark  radii are
  determined by the radii of  the  surrounding clouds.  Quantum
numbers of the constituent quarks are the same as the quantum numbers
of current quarks due to conservation of the corresponding currents
in QCD.

  Collective excitations of the condensate are the Goldstone bosons
and the constituent quarks interact via exchange
of the Goldstone bosons; this interaction is mainly due to pion field.
Pions themselves are the bound states of massive
quarks. The interaction responsible for quark-pion interaction
 can be written in the form \cite{diak}:
 \begin{equation}
{\cal{L}}_I=\bar Q[i\partial\hspace{-2.5mm}/-M\exp(i\gamma_5\pi^A\lambda^A/F_\pi)]Q,\quad \pi^A=\pi,K,\eta.
\end{equation}
The interaction is strong, the corresponding coupling constant is about 4.
The  general form of the total effective Lagrangian (${\cal{L}}_{QCD}\rightarrow {\cal{L}}_{eff}$)
 relevant for
description of the non--perturbative phase of QCD
 includes the three terms \cite{gold} \[
{\cal{L}}_{eff}={\cal{L}}_\chi +{\cal{L}}_I+{\cal{L}}_C.\label{ef} \]
Here ${\cal{L}}_\chi $ is  responsible for the spontaneous
chiral symmetry breaking and turns on first.

To account for the
constituent quark interaction and confinement the terms ${\cal{L}}_I$
and ${\cal{L}}_C$ are introduced.  The  ${\cal{L}}_I$ and
${\cal{L}}_C$ do not affect the internal structure of the constituent
quarks.

The picture of a hadron consisting of constituent quarks embedded
 into quark condensate implies that overlapping and interaction of
peripheral clouds   occur at the first stage of hadron interaction.
The interaction of the condensate clouds assumed to of the shock-wave type,
this condensate clouds interaction generates the quark-pion transient state.
This mechanism is inspired by the shock-wave production process proposed by
Heisenberg \cite{heis} long time ago.
At this stage,  part of the effective lagrangian ${\cal{L}}_C$ is turned off
(it is turned on again in the final stage of the reaction).
Nonlinear field couplings   transform then the kinetic energy to
internal energy \cite{heis,carr}.
As a result the massive
virtual quarks appear in the overlapping region and transient state of matter is generated.
This state consist of $\bar{Q}Q$ pairs and
pions strongly interacting with quarks.

Part of hadron energy carried by
the outer condensate clouds  being released in the overlap region goes to
generation of massive quarks interacting by pion exchange
 and their number was estimated as follows as follows:
\begin{equation} \tilde{N}(s,b)\,\propto
\,\frac{(1-\langle k_Q\rangle)\sqrt{s}}{m_Q}\;D^{h_1}_c\otimes D^{h_2}_c
\equiv N_0(s)D_C(b),
\label{Nsbt}
\end{equation} where $m_Q$ -- constituent quark mass, $\langle k_Q\rangle $ --
average fraction of
hadron  energy carried  by  the constituent valence quarks. Function $D^h_c$
describes condensate distribution inside the hadron $h$ and $b$ is
an impact parameter of the colliding hadrons.
Thus, $\tilde{N}(s,b)$ quarks appear in addition to $N=n_{h_1}+n_{h_2}$
valence quarks.

The generation time of the transient state $\Delta t_{tsg}$ in this picture obeys to the
inequality
\[
\Delta t_{tsg}\ll \Delta t_{int},
\]
where $\Delta t_{int}$ is the total interaction time.
The newly generated massive virtual
quarks play a role of scatterers for the valence quarks in elastic
scattering; those quarks are transient ones in this process: they
are transformed back into the condensates of the final hadrons.

Under construction of the model  for elastic scattering
 \cite{csn}
it was assumed that the valence quarks located in the
central part of a hadron are scattered in a
quasi-independent way off the transient state
 with interaction radius of valence quark  determined
by  its inverse mass:
\begin{equation}\label{rq}
R_Q=\kappa/m_Q.
\end{equation}
The elastic scattering $S$-matrix in the impact parameter representation is
written in the model in the form of linear fractional transform:
\begin{equation}
S(s,b)=\frac{1+iU(s,b)}{1-iU(s,b)}, \label{um}
\end{equation}
where $U(s,b)$ is the generalized reaction matrix, which is considered to be an
input dynamical quantity similar to an input Born amplitude  and
related to the elastic scattering scattering amplitude through an algebraic
equation which enables one to restore unitarity \cite{umat}.
The
function $U(s,b)$  is chosen in the model as a product of the
averaged quark amplitudes \begin{equation} U(s,b) =
\prod^{N}_{Q=1} \langle f_Q(s,b)\rangle \end{equation} in
accordance  with assumed quasi-independent  nature  of the valence
quark scattering.
The essential point here
is the rise with energy of the number of the scatterers  like
$\sqrt{s}$. The $b$--dependence of the function
$\langle f_Q \rangle$  has a simple form $\langle
f_Q(b)\rangle\propto\exp(-m_Qb/\xi )$.

These notions can be extended to particle
production with account of the geometry  of the overlap region and dynamical properties of
the transient state.
Valence constituent quarks would excite a part of the cloud of the virtual massive
quarks and those
quarks will subsequently hadronize  and form the multiparticle
final state.   This mechanism
can be relevant for the region of moderate transverse momenta while the region
of high transverse momenta should be described by the excitation of the constituent
quarks themselves and application of  the perturbative QCD to the parton structure
of the constituent quark.
The model allow to describe elastic scattering
and the main features of multiparticle production \cite{csn,jpg,refl}.
In particular, it leads to asymptotical dependencies
\begin{equation}\label{tota}
  \sigma_{tot,el}\sim \ln^2 s,\;\;
 \sigma_{inel}\sim \ln s, \;\; \bar{n}\sim s^\delta.
\end{equation}

Inclusive cross-section for unpolarized
particles  integrated over impact parameter $\mathbf b $,
cannot depend on the azimuthal angle of the detected particle
transverse momentum.
The $s$--channel unitarity  for the inclusive cross-section  could be
 accounted for  by the following representation
\begin{equation}
\frac{d\sigma}{d\xi}= 4\int_0^\infty
d\mathbf{b}\frac{I(s,\mathbf{b},\xi)}{|1-iU(s,\mathbf{b})|^2}\label{unp}.
\end{equation}
 The set of kinematic variables denoted
by $\xi$ describes the state of the  detected particle.
The function $I$ is constructed from the multiparticle analogs $U_n$ of the function $U$
and is in fact an
non-unitarized inclusive cross-section in the impact parameter space and unitarity corrections
is given by the factor
\[
w(s,b)\equiv |1-iU(s,b)|^{-2}
\]
 in Eq. (\ref{unp}).
Unitarity  modifies anisotropic flows.
When the impact parameter vector $ \mathbf {b}$ and transverse momentum ${\mathbf  p}_\perp $
of the detected particle are fixed,
the function $I=\sum_{n \geq 3} I_n$, where $n$ denotes  a number of particles in the final state,
  depends on the azimuthal angle $\phi$ between
 vectors $ \mathbf b$ and ${\mathbf  p}_\perp $.
It should be noted that the impact parameter
$ \mathbf {b}$ is the  variable conjugated to the transferred momentum
$ \mathbf {q}\equiv \mathbf {p}'_a-\mathbf {p}_a$ between two incident channels
 which describe production processes
of the same final multiparticle state.
The dependence on the azimuthal angle $\phi$ can be written in explicit form through the Fourier
series expansion
\begin{equation}\label{fr}
I(s,\mathbf b, y, {\mathbf  p}_\perp)=I_0(s,b,y,p_\perp)[1+
\sum_{n=1}^\infty 2\bar v_n(s,b,y,p_\perp)\cos n\phi].
\end{equation}
The function $I_0(s,b,\xi)$ satisfies  to the
following sum rule
\begin{equation}\label{sumrule}
\int I_0(s,b,y,p_\perp) p_\perp d p_\perp dy=\bar n(s,b)\mbox{Im} U(s,b),
\end{equation}
where $\bar n(s,b)$ is the mean multiplicity depending on impact parameter.
The ``bare''  flow $\bar v_n$ is related to the
measured  flow $v_n$  as follows
\[
v_n(s,b,y,p_\perp)=w(s,b)\bar v_n(s,b,y,p_\perp).
\]
In the above formulas the variable $y$ denotes rapidity, i.e. $y=\sinh^{-1}(p/m)$,
where $p$ is a longitudinal momentum.
Evidently, corrections due to unitarity are mostly important
at small impact parameters, i.e. they provide an additional suppression of the
 anisotropic flows at small centralities,
while very peripheral collisions are not affected by  these corrections.

The  geometrical picture of hadron collision at non-zero impact parameters
described above
implies that the generated massive
virtual  quarks in overlap region will obtain large initial orbital angular momentum
at high energies. The total orbital angular
momentum  can be estimated
as follows
\begin{equation}\label{l}
 L(s,b) \simeq \alpha b \frac{\sqrt{s}}{2}D_C(b).
\end{equation}
The parameter $\alpha$ is related to the fraction of the initial energy carried by the condensate
clouds which goes to rotation of the quark system and
the overlap region, which is described by the function $D_C(b)$,
has an ellipsoidal form (Fig. 1).
\begin{figure}[hbt]
\begin{center}
\epsfxsize=  70 mm  \epsfbox{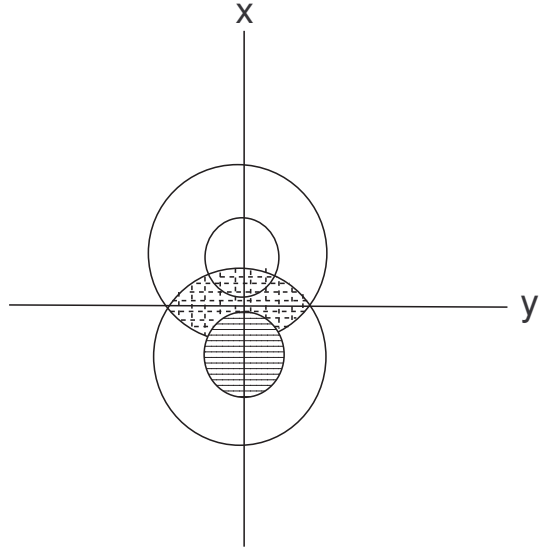}
\end{center}
\caption{Schematic view in frontal plane of the hadron collision as extended objects.
Collision occurs along the z-axis.}
\end{figure}
It should be noted that $L\to 0$ at $b\to\infty$ and $L=0$ at $b=0$ (Fig. 2).
\begin{figure}[hbt]
\begin{center}
\epsfxsize=  70 mm  \epsfbox{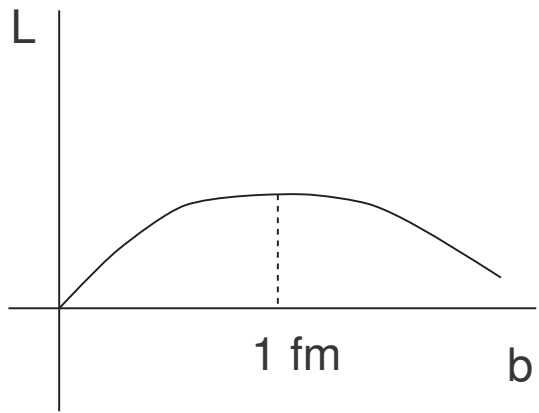}
\end{center}
\caption{Qualitative dependence of the orbital angular momentum $L$ on the impact parameter $b$.}
\end{figure}

\section{Rotating transient matter and directed flow \\ in hadronic reactions}
Now we would like to discuss the experimental consequences of the described picture of hadron
collisions. The important problem here is the experimental determination of the impact parameter
$\mathbf b$.
To proceed that way  the measurements of the characteristics of
 multiparticle production processes in hadronic collisions
at fixed impact parameter should be performed with  selection of the
specific events sensitive to the value and direction of impact parameter.
The
determination of the reaction plane in the non-central hadronic collisions could be
 experimentally realizable   with the utilization of the standard procedure\cite{polvol,poskanz}.
The relationship of the impact parameter with the
 final state multiplicity is a useful tool in these studies similar to the studies
 of the nuclei interactions.  For example, in the Chou-Yang geometrical approach  \cite{chyn}
  one can restore the
 values of impact parameter from the charged particle multiplicity
\cite{chmult}.
The centrality is determined by the fraction of the events with the largest number of
 produced particles which are registered by detectors (cf. \cite{antinori}).
Thus, the impact parameter can be  determined through
 the centrality and then, e.g. directed flow, can be analyzed by selecting events
 in a specific centrality ranges.
Indeed,  the relation
 \begin{equation}\label{cent}
 c(N)\simeq \frac{\pi b^2(N)}{\sigma_{inel}},
\end{equation}
between centrality and impact parameter was obtained \cite{bron} and can be
 extended straightforwardly to the case of hadron scattering.
 In this case we should consider $\bar{R}$ as a sum of the two radii of colliding hadrons
 and $\sigma_{inel}$ as the total inelastic hadron-hadron cross--section. The centrality
 $c(N)$ is the centrality of the events with the multiplicity larger than $N$ and $b(N)$ is
 the impact parameter where the mean multiplicity $\bar n (b)$ is equal to $N$.

At this point we would like to make an important remark. Namely, based on the recent
discoveries at RHIC, we assume  that the matter in
the transient (intermidiate) state of hadronic reactions has a strongly interacting nature.
Due to strong interaction
between quarks in the transient state, it can be described as  a liquid.
Therefore, the orbital angular momentum $L$ should be realized  as a coherent rotation
of the quark-pion liquid  as a whole  in the
$xz$-plane
(due to mentioned strong correlations between particles presented in the liquid). It should
be noted that for the given value of the orbital angular momentum $L$
 kinetic energy has a minimal value if all parts of liquid rotates with the same angular velocity.
  We  assume therefore
that the different parts of the quark-pion liquid in the overlap region indeed have the same
angular velocity $\omega$.
In this model spin of the polarized hadrons has its origin in the
rotation of matter hadrons consist of.
In contrast, we assume rotation of the matter
during intermediate, transient state of hadronic interaction.

Collective rotation
of the strongly interacting system of the massive
constituent quarks and pions is  the main point of the proposed
  mechanism of the directed flow generation in hadronic and nuclei collisions.
We  concentrate on the effects of this rotation and consider directed flow for the constituent quarks
supposing that  directed flow for  hadrons is close to the directed
flow for the constituent quarks at least qualitatively.

The assumed particle production mechanism at moderate transverse
momenta is
an excitation of  a part of the rotating transient state of  massive constituent
quarks (interacting by pion exchanges) by the one of the valence constituent quarks
 with  subsequent hadronization
of the quark-pion liquid droplets.
Due to the fact that the transient matter is strongly interacting, the excited parts
should be located closely  to the periphery  of the rotating transient state otherwise absorption
 would not allow
to quarks and pions to leave the region (quenching). The mechanism is sensitive
 to the particular
rotation direction and the directed flow should have  opposite signs for the particles
in the fragmentation regions of the projectile and target respectively.
It is evident that the effect of  rotation (shift
in  $p_x$ value ) is
most significant in the peripheral part of the rotating quark-pion liquid
and is to be weaker in the less peripheral regions (rotation with the same angular velocity $\omega$),
i.e. the directed flow $v_1$ (averaged over all transverse
momenta)  should be proportional to the inverse depth $\Delta l$
where the excitation of the rotating quark-pion liquid takes place (Fig. 3)
\begin{figure}[hbt]
\begin{center}
\epsfxsize=  70 mm  \epsfbox{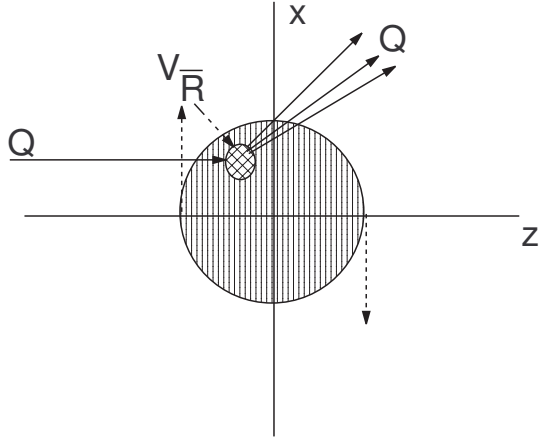}
\end{center}
\caption{Interaction of the constituent quark with rotating quark-pion liquid.}
\end{figure}
i.e.
\begin{equation}\label{v1}
|v_1|\sim \frac{1}{\Delta l}
\end{equation}
In its turn, the length   $\Delta l$ is related to the energy loss
of constituent valence quark in the medium due to elastic
rescattering (quark-pion liquid) prior an excitation occurs, i.e.
before constituent quark would deposit its energy into the energy
of the excited quarks (those quarks lead to the production of the
secondary particles), i.e. it can be  assumed
that
\[
\Delta l\sim \Delta y,
\]
where $\Delta y =|y-y_{beam}|$ is the
difference between the rapidities  of the final particle and  the projectile.
On the other hand, the  depth length $\Delta l$  is determined
  by elastic quark scattering cross-section $\sigma$ and quark-pion liquid density $n$.
Therefore, the averaged value of $v_1$ should be proportional to the particle density of
the transient state and include  cross-section $\sigma $, i.e.
   \begin{equation}\label{v1dep}
   \langle |v_1|\rangle \sim {\sigma n}.
   \end{equation}
   This estimate shows that the magnitude of the directed flow could provide information
   on the properties of the transient state.
   The magnitude of observable $v_1$  (determined by the shift of transverse momentum
due to rotation) is proportional to $(\Delta l)^{-1}$ in this mechanism and
 depends on  the rapidity difference as
 \begin{equation}\label{v1yd}
 |v_1|\sim
\frac{1}{|y-y_{beam}|}
\end{equation}
  and does not depend on the incident energy.   Evidently,
  the directed flow $|v_1|$ decreases
  when the absolute value of the above difference increases, i.e. $|v_1|$ increases at
  fixed energy and increasing rapidity of final particle and it decreases at fixed rapidity
  of final particle and increasing  beam energy. Dependence of $|v_1|$ will be
   universal for different energies  when plotted against the difference $|y-y_{beam}|$ (Fig.4).
\begin{figure}[hbt]
\begin{center}
\epsfxsize=  70 mm  \epsfbox{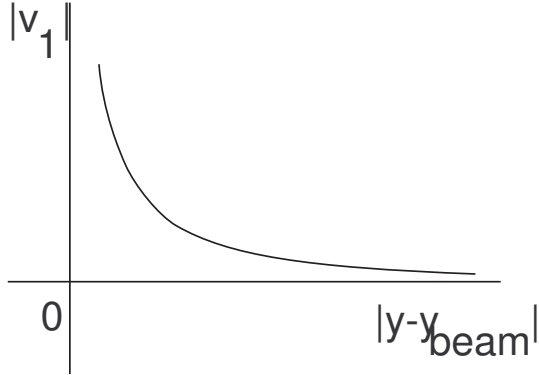}
\end{center}
\caption{Universal dependence of directed flow on the rapidity difference.}
\end{figure}

  The centrality dependence of $|v_1|$ is determined by the orbital momentum dependence
  $L$ on the impact parameter, i.e. it should   be decreasing towards  high and lower centralities.
  Decrease toward high centralities is evident, no overlap of hadrons or nuclei should be at high enough
  impact parameters. Decrease of $v_1$ toward lower centralities is specific prediction of the proposed
  mechanism based on rotation
  since central collisions with smaller impact parameters would lead to slower rotation or its complete
   absence in the head-on collisions\footnote{Of course, there is another reason for vanishing $v_1$ at
   $b=0$, it is rotational invariance around collision axis.}.
  Thus,  the qualitative centrality dependence of $|v_1|$ corresponds to Fig. 2.

 Now we  can consider
transverse momentum dependence of the directed flow $v_1(p_\perp)$ (integrated over rapidity)
for constituent quarks. It is natural to
suppose that the size of the region where the virtual massive quark $Q$
comes from the quark-pion liquid
is determined by its transverse momentum, i.e. $\bar R\simeq 1/p_\perp$. However, it is
evident that $\bar R$ should not be larger than the interaction radius of the valence
 constituent quark $R_Q$ (interacting
with the
  quarks and pions from the transient liquid state). The production processes with
high transverse momentum such that   $\bar R$ is much less than the
geometrical size of the valence constituent quark $r_Q$ resolve
its internal structure as a cluster of the
non-interacting partons.  Thus,  at high transverse momenta the constituent quarks will be excited
themselves and  hadronization of the uncorrelated partons would lead to the secondary
particles with high transverse momenta and  vanishing directed flow.  If the production mechanism
rendering to the constituent quark excitation is valid, then  similar
 conclusions on the small anisotropic flows at large transverse momenta should be applicable for
 $v_n(p_\perp)$  with $n>1$. Obviously, it should not be valid in the case of the polarized hadron collisions.

 The magnitude of the quark
interaction radius $R_Q$ can be taken from
 the analysis of elastic scattering \cite{csn};
 it has dependence on its  mass in the form (\ref{rq}) with
$\kappa \simeq 2$, i.e. $R_Q\simeq 1$ $fm$ \footnote{This is the light constituent quark
interaction radius which is close to the inverse
pion mass.}, while the geometrical radius of  quark $r_Q$
is about $0.2$ $fm$.
The size of the region\footnote{For simplicity we suppose that this region has a spherically
symmetrical form}, which is responsible for the small-$p_\perp$ hadron production, is large,
valence constituent quark excites rotating cloud of quarks with various values and directions
of their momenta in that case. Effect of rotation will be averaged over  the volume $V_{\bar R}$
 and therefore $\langle \Delta p_x \rangle_{V_{\bar R}}$ and $v_1(p_\perp)$ should be small.

When we proceed to the region of higher values of $p^Q_\perp$, the radius $\bar R$ is decreasing
and the  effect of rotation  becomes  more prominent, valence quark excites now the region
where most of the quarks move coherently in the same direction with approximately
equal velocities.
 The mean value $\langle \Delta p_x \rangle_{V_{\bar R}}$
and the directed flow, respectively,
can have a significant magnitude and increase with increasing $p_\perp$.
When $\bar R$ becomes smaller than the geometrical radius
of constituent quark, the interactions at short distances start to resolve its internal
 structure as an uncorrelated cloud of partons.
 The production of the hadrons at such high values of transverse momenta is due to the
excitation of the constituent quarks themselves and  subsequent hadronization
of the partons.
The collective effects of rotating transient cloud in $v_1$ at large $p^Q_\perp > 1/r_Q$
will disappear as well as the directed flow of final particles.
The value of transverse
momentum, where the maximal values in the $p_\perp$-dependence of $v_1$ are expected,
are in the region $1$ $GeV/c$ since $r_Q\simeq 0.2$ $fm$. The qualitative $p_\perp$-dependence
of the directed flow is illustrated on Fig. 5.
\begin{figure}[hbt]
\begin{center}
\epsfxsize=  70 mm  \epsfbox{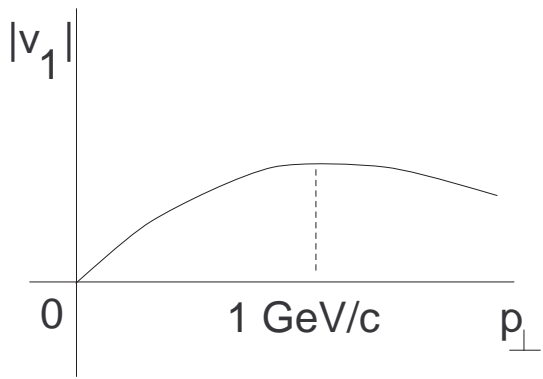}
\end{center}
\caption{Qualitative dependence of directed flow on transverse momentum $p_\perp$.}
\end{figure}

\section{Directed flow in nucleus collisions}

Until now we considered hadron scattering and directed flow in this process, but
significant ingredient in this consideration was borrowed from the $AA$ studies, namely
we supposed that the  transient matter is a strongly interacting one and it is the same in
$pp$ and $AA$ reactions.  Using the experimental findings of RHIC, we assumed,in fact,
 that it is a  liquid consisting of massive quarks, interacting by pion exchange, and
  characterized  by the fixed interparticle distances determined by the quark interaction radius.
  The assumption on the almost instantaneous, shock-wave type of
generation of the transient state obtains then  support in the very short thermalization time revealed
in heavy-ion collisions at RHIC \cite{therm}. Existence of the massive quark matter in the stage
preceding hadronization seems to be supported also by the experimental data obtained
at CERN SPS \cite{biro}.

The geometrical picture of hadron collision has an apparent analogy
with collisions of nuclei and
it should be noted that the appearance of large orbital angular momentum
should be expected in the overlap region in the non-central nuclei collisions.
 And then due to strongly interacting
nature of the transient matter we assume that this orbital angular momentum realized
as a coherent rotation of liquid. Thus, it seems that  underlying  dynamics  could be   similar
to the dynamics of the directed flow in
hadron collisions.

We can go further and extend the production mechanism from hadron to
nucleus case also. This extension cannot be straightforward. First, there
will be no unitarity corrections for the anisotropic flows and instead
of valence constituent quarks, as a projectile we should consider nucleons,
which would excite rotating quark liquid. Of course, those differences will
result in significantly higher values of directed flow. But, the general trends in its
dependence on the collision energy, rapidity of the detected particle and transverse momentum,
should be the same. In particular, the directed flow in nuclei collisions as well as
in hadron reactions will depend
on  the rapidity difference $y-y_{beam}$ and not on the incident energy.
The mechanism
 therefore can provide a qualitative explanation of the incident-energy scaling
 of $v_1$ observed at RHIC \cite{v1exp}.
  In the
projectile frame the directed flow has the same values for the  different
 initial energies. In the  Fig. 6 experimental data are shown along with the dependence
 \begin{equation}\label{vv}
 v_1\sim (\eta-y_{beam})^{-1},
 \end{equation}
 where $\eta$ is a pseudorapidity. This dependence reflects  the trend  in the
 experimental data and, therefore, the mechanism described in the previous section obtains
 an experimental justification, qualitative, of course.
\begin{figure}[h]
\begin{center}
  \resizebox{8cm}{!}{\includegraphics*{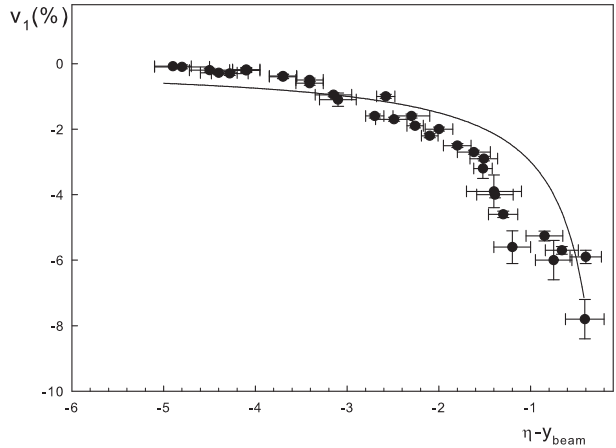}}
\end{center}
\caption{Dependence of directed flow on $\eta-y_{beam}$ in $Au+Au$ and $Cu+Cu$ collisions
at 62.4 GeV and 200 GeV at RHIC (preliminary data of STAR Collaboration \cite{v1exp})}
\end{figure}
In addition, we would like to note that azimuthal dependence of the suppression factor $R_{AA}$
found in the experiment PHENIX at RHIC \cite{pant} can also be explained by rotation of the transient state.
Since the correlations result from the rotation in this mechanism and therefore are maximal
in the rotation plane, similar dependence should be observed in the azimuthal dependence
of the two-particle correlation function.
Effect of rotation should be maximal for the peripheral collisions and therefore the
dependence on $\phi$ should be most steep at larger   impact parameter (or centrality) values
The discussed rotation mechanism  should  contribute to the elliptic flow too.
However, since the regularities already found experimentally for
  $v_1$ and $v_2$ in nuclei interactions imply  different dynamical origin for these flows, we should conclude
  that this mechanism does not provide  significant contribution to the elliptic flow.

\section{Directed flow at the LHC energies and conclusion}

Nowadays, with approaching start of the LHC, it is interesting to predict  what should be
expected at such very high energies \cite{last}, in particular would deconfined matter produced in
$pp$ and $AA$ collisions be weakly interacting or it will remain to be a strongly interacting one
as it was observed at RHIC in $AA$ collisions? In the latter case one can expect that proposed
rotating mechanism will be working at the LHC energies and therefore  the observed at RHIC incident-energy
scaling in $v_1$ will remain to be valid also,  i.e. directed flow plotted against the difference
$y-y_{beam}$ will be the same as it is at RHIC.
However, if the transient matter at the LHC energies will be weakly interacting, then one should
expect absence of the coherent rotation and the vanishing of the directed flow.
This conclusion is valid provided that the rotation is the only mechanism of the directed flow
generation. If it is so, vanishing directed flow can serve as a  signal of a genuine quark-gluon plasma
(gas of free quarks and gluons) formation.

The parton interaction and finite transverse gradient of parton
longitudinal momentum  is a driving force of the orbital angular
momentum conversion to the global system polarization through
spin-orbital coupling \cite{wang}. In the parton picture,which is a weakly
interacting medium, the coherent rotation is absent. The predicted global  polarization
was not found  experimentally at RHIC \cite{starpol} and it is just
another, indirect confirmation of the strongly interacting character
of the deconfined matter at such energies. It however does not mean that at the
LHC energies  picture will be the same.
Thus, if the matter produced at the LHC is weakly interacting and since the total
angular momentum should be conserved in any case, the orbital momentum would be converted
then into global polarization at the partonic
level which can be detected experimentally  measuring hyperon or photon
polarizations \cite{wang, ipp}.
It should be noted that there is another reason
for the emerging global polarization of the secondary particles in hadron collisions at the LHC
energies --- it is antishadowing or reflective scattering \cite{refl,imb}. The latter mechanism is not
related to the question of what kind of deconfined matter was produced.

Thus, it would be interesting to perform the measurements of the
anisotropic flows at RHIC  and at the LHC not only in heavy ion
collisions, but  in $pp$--collisions also, and to find possible
existence or absence of the rotation effects related to strong
interactions through the directed flow studies. At  RHIC there is
another possibility of the directed flow measurements in the
polarized proton collisions. Those studies when combined with the
measurements of the directed flow in the unpolarized
$pp$--scattering, could provide information on the  role of the
orbital angular momentum in the spin structure of the nucleon.
At large transverse momenta, they will provide information on the
generalized parton distributions in the impact parameter space,
which are under active studies nowadays (cf. \cite{burk}).

We discussed  qualitative  features of the transient state in hadronic
and nuclei collisions related to the directed flow $v_1$.  We concentrated on the
hadron interactions. However,  the most conclusions should remain to be valid
for nucleus interactions.

\section*{Acknowledgement}
We are grateful to J. Dunlop, V. Pantuev and O. Teryaev for the interesting discussions.
We would like to acknowledge also valuable remarks and comments from the Referees and
the Editorial board.

\end{document}